\documentclass[prl,twocolumn,superscriptaddress,floatfix]{revtex4}
\usepackage{graphicx,amsfonts,amssymb,amsmath,hyperref}

\newif\ifhyper
\hypertrue
\ifhyper
\hypersetup{
   citecolor = {green},
   colorlinks = {true}, 
   urlcolor = {blue} 
}
\fi

\begin{document}

\title{Comment on ``Anyonic braiding in optical lattices"}

\author{Julien Vidal}
\email{vidal@lptmc.jussieu.fr}
\affiliation{Laboratoire de Physique Th\'eorique de la Mati\`ere Condens\'ee,
  CNRS UMR 7600, Universit\'e Pierre et Marie Curie, 4 Place Jussieu, 75252
  Paris Cedex 05, France}

\author{S\'ebastien Dusuel}
\email{sdusuel@gmail.com}
\affiliation{Lyc\'ee Louis Thuillier, 70 Boulevard de Saint Quentin,
  80098 Amiens Cedex 3, France}

\author{Kai Phillip Schmidt}
\email{schmidt@fkt.physik.uni-dortmund.de}
\affiliation{Lehrstuhl f\"ur theoretische Physik, Otto-Hahn-Stra\ss e 4, D-44221
Dortmund, Germany}

\maketitle

In Ref.~\cite{Zhang07}, C. Zhang {\it et al.} proposed an experimental scheme to detect the braiding statistics in Kitaev's honeycomb model \cite{Kitaev06}. 
Although the perspective to observe anyons experimentally is very challenging, we would like to point out some major technical and conceptual mistakes which invalidate the conclusion drawn in \cite{Zhang07}. 

Indeed, the study presented in \cite{Zhang07} relies on a perturbation theory of the Kitaev model in the gapped phase ($J_z \gg J_x,J_y$) \cite{Kitaev06,Schmidt08}. At order 0, the low-energy eigenstates are built from collinear dimer configuration on each $z$-dimer. However, contrary to what is assumed in \cite{Zhang07}, this is no longer true when switching the perturbation on, 
as can be seen for instance in a first order calculation of
the ground state correlation function $\langle \sigma^x_i \sigma^x_j \rangle=\frac{J_x}{2J_z}$ on a $x-$dimer $(i,j)$.
This basic error has led the authors to propose a form of the ground state which gives 
$\langle \sigma^x_i \sigma^x_j \rangle\neq 0$ on a $z$-dimer $(i,j)$ whereas exact results show that all two-spin correlations vanish except
$\langle \sigma^\alpha_i \sigma^\alpha_j \rangle$ on a $\alpha$-dimer \cite{Baskaran07}.

In addition, all spin operations used in \cite{Zhang07} to create anyons do also produce fermionic excitations that are likely to spoil the proposed detection set-up \cite{Dusuel08}. Once again, we emphasize that this effect already occurs at first order in perturbation. 

To summarize, results given in \cite{Zhang07} are only valid for $J_x=J_y=0$ where the model is trivial and all low-energy levels exactly degenerate. In other words, the proposal of measuring anyonic excitations suggested in \cite{Zhang07} is strongly questionable.


\end{document}